\documentclass[12pt]{article}
\usepackage{amssymb,amsmath}
\usepackage{cite}
\newlength{\dinwidth}
\newlength{\dinmargin}
\def\idty{{\leavevmode\hbox{\rm 1\kern -.3em I}}}

\def\Hs{{\cal H}}

\def\idty{{\leavevmode\hbox{\rm 1\kern -.3em I}}}

\def\RR{{\mathbb R}}

\def\NN{{\mathbb N}}
\def\IN{{\mathbb N}}

\def\beq{\begin{equation}}
\def\eeq{\end{equation}}
\begin{document}
\title{``On the Impossibility of a Poincar\'e--Invariant \\
Vacuum State with Unit Norm'' \\
Refuted}
\author{{\Large Stephen J.\ Summers\,}\\[5mm]
Department of Mathematics \\
University of Florida \\ Gainesville FL 32611, USA}

\date{February 26, 2008}

\maketitle 

{\abstract Assertions made in a document recently deposited in the arXiv 
are refuted.}

     Recently, a literally incredible document was deposited in
the arXiv \cite{idiot}. I shall restrict my comments to a direct 
rebuttal of the author's central point, which
is indicated in the title of the document. The author claims
``The axioms of QFT cannot be made internally consistent.''
This conclusion is arrived at on the grounds that ``It is impossible
for any state to be both Poincar\'e--invariant and also have
unit norm.''

     For most readers, it will suffice to be reminded that models
satisfying ``the axioms of QFT''\footnote{From the context, the author
is referring to the Wightman axioms \cite{StWi}.} have been
constructed with full mathematical rigor by a number of techniques and
approaches \cite{Ar,BLT,BrGuLo,GlJa,GlJa2,ReSi,WiGa}.\footnote{This list
of references is far from exhaustive.} In these models the state which
the author claims cannot exist {\it does}, in fact, exist. However, if
there is anyone remaining who doubts that the author's reasoning must be
faulty, I shall point out some of the errors in his ``proof'' that ``
It is impossible for any state to be both Poincar\'e--invariant and
also have unit norm.'' These errors obviate his conclusion.

     For the reader's convenience, I shall review the author's
argument. However, I shall use more standard and consistent notation
where necessary. He begins with a separable and infinite dimensional
Hilbert space $\Hs$ and an unspecified orthonormal basis 
$\{ \Psi_n \}_{n \in \NN}$ for $\Hs$. He takes a vector $\Psi \in \Hs$
and expands it with respect to the given basis:
\begin{equation} \label{1}
\Psi = \sum_{n = 1}^\infty c_n \Psi_n \, .
\end{equation}
Then the problems begin.

     To proceed, a few noncontroversial clarifying remarks will be
useful. The translation subgroup of the isometry group of four 
dimensional Minkowski space is usually realized
as the additive group of four dimensional real vectors, denoted here
by $\RR^4$. For well known, physically motivated reasons, this 
symmetry group acts upon the Hilbert space of states solely through the 
intermediary of a unitary representation $U(\RR^4)$ \cite{StWi,BLT,Ba,We}.  
Hence, the phrase ``a vector $\Psi \in \Hs$ is translation--invariant'' 
has the following mathematical meaning:
\begin{equation} \label{2}
U(a) \Psi = \Psi \, ,
\end{equation}
for all $a \in \RR^4$. 

     Here comes the rub: the author, without a word of justification,
claims the following is true:
\begin{equation} \label{b1}
U(a) \sum_{n = 1}^\infty c_n \Psi_n = \sum_{n = 1}^\infty c_{n+a} \Psi_n \, ,
\end{equation}
for all $a \in \RR^4$. Then, for a translation--invariant vector it
follows from equations ({\ref{1}), (\ref{2}) and 
(\ref{b1}) that
$$\sum_{n = 1}^\infty c_n \Psi_n = \sum_{n = 1}^\infty c_{n+a} \Psi_n \, ,$$
and the orthonormality of the basis then yields
\begin{equation} \label{b2}
c_n = c_{n+a} \, ,
\end{equation}
for all $n \in \IN$ (in particular, for $n = 1$) and $a \in \RR^4$. He 
then (tacitly) lets $a$ run through the natural numbers (!!!) to conclude 
$c_1 = c_n$ for all $n \in \IN$. Of course, since for any vector 
$\Psi \in \Hs$ one must have $\lim_{n \rightarrow \infty} c_n = 0$, 
this yields $\Psi = 0$.\footnote{Note that $\Psi \in \Hs$ entails 
$\Vert \Psi \Vert$ must be finite.} The author therefore ``proves''
that the only translation--invariant vector in $\Hs$ is the zero vector.

     Doubtless, most readers have recognized equation (\ref{b1})
and the argument after (\ref{b2}) to be nonsensical, but just in 
case someone misses the point, I shall explain.

     (1) Even if the equation made sense, (\ref{b1}) does not
represent the action of the translation group in any known quantum field
model --- cf. any of the examples cited above --- and therefore it is
certainly not entailed by the Wightman axioms. To emphasize the
point: equation (\ref{b1}) is purely the author's fantasy, which
he does not even try to support.

     (2) However, equation (\ref{b1}) is not even mathematically 
meaningful. Since $n \in \IN$ and $a \in \RR^4$, what does $n + a$ mean?
What does $c_{n + a}$ mean if $n + a$ is not a natural number? Has
it become necessary to point out the fact that if one allows meaningless
quantities into an argument, then {\it anything} can be ``proven'',
including $1 = 2$?
     
     (3) In addition, I point out the following facts: The mapping
$$V \sum_{n = 1}^\infty c_n \Psi_n = \sum_{n = 1}^\infty c_{n+k} \Psi_n$$
(which {\it does} make sense when $k \in \IN$ and to which\footnote{This
operator is the $k$th power of the adjoint of the unilateral shift
operator well known to functional analysts, and it has long been known
that the powers of this adjoint tend to the zero operator in the strong
operator topology, cf. \cite{Fi,Ni}.} the author 
is tacitly appealing when he employs the argument after 
(\ref{b2})) depends upon the choice of the basis 
$\{ \Psi_n \}_{n \in \NN}$ (of which there are uncountably infinitely
many in a separable Hilbert space) and is not unitary, 
for any choice of $k \in \IN$. For these and many other reasons which
need not be detailed here, such operators have no physical interpretation
in quantum field theory. And, what is more to the point, they have nothing 
to do with the spacetime symmetry translations the Wightman axioms 
actually refer to.


\begin{thebibliography}{Bor}
\footnotesize

\bibitem{Ar}
H. Araki, Von Neumann algebras of local observables for free scalar
field, {\sl J. Math. Phys., \bf 5}, 1--13 (1964).

\bibitem{Ba}
V. Bargmann, Note on Wigner's theorem on symmetry operations, 
{\sl J. Math. Phys., \bf 5}, 862--868 (1964).

\bibitem{idiot}
J. Berkowitz, On the impossibility of a Poincar\'e--invariant
vacuum state with unit norm, preprint arXiv:0802.0216.

\bibitem{BLT}
N.N. Bogolubov, A.A. Logunov and I.T. Todorov, {\it Introduction to 
Axiomatic Quantum Field Theory}, (W.A. Benjamin, Reading, Mass.) 1975
(translation of Russian original, published in 1969).

\bibitem{BrGuLo}
R. Brunetti, D. Guido and R. Longo, Modular localization and Wigner
particles, {\sl Rev. Math. Phys., \bf 14}, 759--785 (2002).

\bibitem{Fi}
P.A. Fillmore, The shift operator, {\sl Amer. Math. Monthly, \bf 81},
717--723 (1974).

\bibitem{GlJa}
J. Glimm and A. Jaffe, Boson quantum field theory models, in:
{\it Mathematics of Contemporary Physics}, edited by R.F. Streater, 
(Academic Press, London) 1972.

\bibitem{GlJa2}
J. Glimm and A. Jaffe, {\it Quantum Physics}, (Springer Verlag, Berlin)
1981.

\bibitem{Ni}
N.K. Nikol'skii, {\it Treatise on the Shift Operator, Spectral Function
Theory}, (Springer Verlag, Berlin, Heidelberg, New York and Tokyo) 1986. 

\bibitem{ReSi}
M. Reed and B. Simon, {\it Methods of Modern Mathematical Physics,
II: Fourier Analysis and Self--Adjointness}, (Academic Press,
New York, San Francisco and London) 1975.

\bibitem{StWi}
R.F. Streater and A.S. Wightman, {\it PCT, Spin and Statistics, and
All That}, (Benjamin/Cummings Publ. Co., Reading, Mass.) 1964;
second printing, with additions and corrections, 1978.

\bibitem{We}
S. Weinberg, {\it The Quantum Theory of Fields}, Vol. 1 (Cambridge 
University Press, Cambridge) 1995.

\bibitem{WiGa}
A.S. Wightman and L. G\aa rding, Fields as operator--valued distributions
in relativistic quantum field theory, {\sl Ark. f. Fys., \bf 28}, 129--184
(1964).

\end{thebibliography}
\end{document}